\documentclass{webofc}

\usepackage[varg]{txfonts}   
\usepackage{hyperref}
\usepackage{url}
\hypersetup{colorlinks=true,citecolor=blue,urlcolor=blue,linkcolor=blue}
\begin{document}
\title{Enhanced production of charged over neutral kaons 
in~Ar+Sc collisions measured by NA61/SHINE at CERN SPS}

\author{
\firstname{Tomasz} \lastname{Matulewicz}
\inst{1}\fnsep\thanks{\email{tomek.matulewicz@cern.ch}} 
for the NA61/SHINE Collaboration
}

\institute{Faculty of Physics, University of Warsaw, 02-093 Warsaw, Poland}

\abstract{The isospin symmetry, originating from similar  masses of $u$ and $d$ quarks, if exact would result in equal numbers of charged ($K^+$ and $K^{-}$) and neutral ($K^0$ and $\overline{K}^0$) mesons produced in collisions of isospin-symmetric atomic nuclei. 
The charged and neutral $K$ meson production in Ar+Sc  collisions at a center-of-mass energy of 11.9 GeV per nucleon pair was measured by the NA61/SHINE Collaboration. 
The production of charged $K$ mesons at mid-rapidity is (18.4$\pm$6.1)\% higher than that of the neutral $K$ mesons.
The models of hadron production, including known isospin-symmetry breaking effects, cannot explain the measurements.}

\maketitle
\section{Introduction}
\label{intro}
The mass of strange mesons $K$ belonging to two isospin doublets ($K^+$, $K^0$) and ($K^-$, $\overline{K}^0$) differs by only $\sim$ 1\%, so the production of these particles in isospin-symmetric collision systems can be expected to be very similar.
The charged kaons can be directly measured and identified.
The neutral $K$ mesons oscillate and they are observed as $K_S^0$ and $K_L^0$ objects of very different lifetime.
The $K_S^0$ meson decays (69.2\%) into a pair of charged pions.
The reconstruction of these two-body decays, originating from a secondary vertex, allows to determine the yield of $K_S^0$ mesons by the invariant mass analysis.
The yield of $K_S^0$  is then expected to be equal to the average of charged ones, provided an isospin-symmetric collision system as a source
\begin{equation*}
    R_K=\frac{\langle K^+\rangle+\langle K^-\rangle}{\langle K^0\rangle+\langle\overline{K}^0\rangle}=\frac{\langle K^+\rangle+\langle K^-\rangle}{2\langle K^0_S\rangle}=1.
\end{equation*}

\section{Experiment}
\label{experiment}
The NA61/SHINE is a fixed-target detector designed to study the hadron production in hadron-proton, hadron-nucleus, and nucleus-nucleus interactions at $\sqrt{s_{NN}}=5-17$ GeV. 
It is situated in the North Area of the SPS accelerator at CERN. 
The spectrometer covers nearly the entire forward hemisphere for charged and neutral (decaying to charged) hadrons. 
Some neutral hadrons are detectable below mid-rapidity.

The main devices for hadron measurement are four large Time Projection Chamber detectors, two of them placed in magnetic field. 
The identification of slower particles is supplemented by Time of Flight measurement. 
The centrality of the collision is determined by the energy deposited in the forward calorimeter.
The detector has been recently upgraded, resulting in the increase of event rate capability by more than factor 10.

\section{Results}
\label{results}

\begin{figure}
\centering
\sidecaption
\includegraphics[width=9cm,clip]{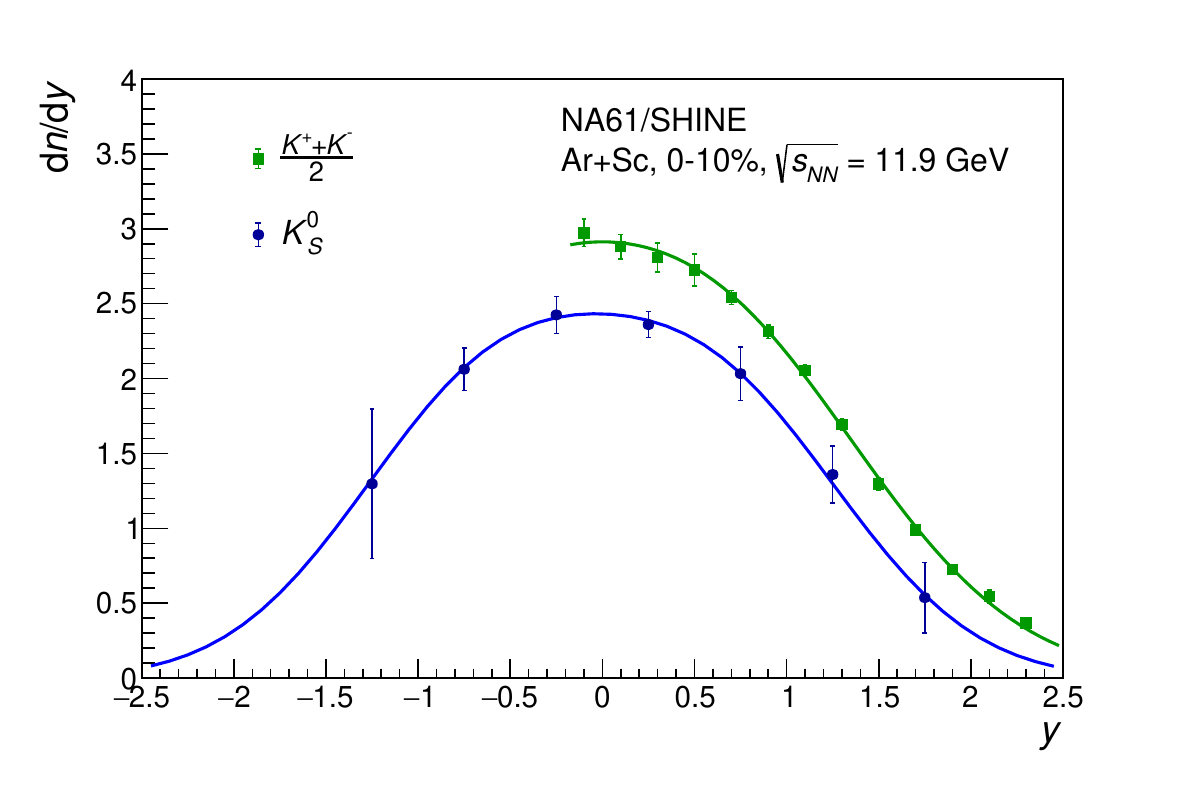}
\caption{Rapidity distributions of charged ($K^+$+$K^-$)/2 and $K^0_S$ mesons measured in 10\% central Ar+Sc collisions at $\sqrt{s_{NN}}$=11.9 GeV \cite{NA61nature}.}
\label{fig_dndy}      
\end{figure}

The double-differential ($y$, $p_T$) yields of $K^+$, $K^-$, and $K^0_S$ mesons were measured in the 10\% most central collisions at beam momentum 75$A$ GeV/$c$ in Ar+Sc system.
Integrating the $p_T$ distributions, the rapidity distributions were obtained (Fig. \ref{fig_dndy}),  showing the overproduction of charged kaons.
The $R_K$ ratio is above 1 for numerous measurements \cite{NA61nature}, but only the NA61/SHINE result (at $\sqrt{s_{NN}}=11.9$ GeV) is 3$\sigma$ significant.
Several isospin-violating effects (non-zero isospin in the entrance channel, different mass and charge of $u$ and $d$ quarks, unequal contribution from the $\phi(1020)$-meson decay) are accounted for in hadron-resonance gas (HRG) model and are unable to explain the observed deviation \cite{NA61nature}. 
The preliminary result at $\sqrt{s_{NN}}$=8.8 GeV (Table \ref{tab:RK}) is consistent with the published value at $\sqrt{s_{NN}}$=11.9 GeV. 
Results of analysis at $\sqrt{s_{NN}}$=16.8 GeV are expected soon.
The experiments with isospin-symmetric systems oxygen-oxygen and magnesium-magnesium are planned to better elucidate the effect (42 million oxygen-oxygen events were already measured at $\sqrt{s_{NN}}$=16.8 GeV during summer 2025).

\begin{table}[htb]
\centering
\caption{Values of $R_K$ ratio in Ar+Sc collisions (two upper rows) and the average over measurements calculated from kaon yields reported in the literature (bottom row, mainly Au+Au collisions).}
\begin{tabular}{ccc}
 $\sqrt{s_{NN}}$ (GeV) & $R_K$  &  \\ \hline\hline
8.8 & 1.115 $\pm$ 0.043 & NA61/SHINE preliminary    \\ \hline
11.9 & 1.184 $\pm$ 0.061 & Ref. \cite{NA61nature}   \\ \hline 
7.6 -- 2760 & 1.129 $\pm$ 0.027 & average \cite{NA61nature}
\end{tabular}
    \label{tab:RK}
\end{table}

This project was co-financed by the Ministry of Science and Higher Education of the Republic of Poland within the framework of the "Excellent Science – Integrated Development" program (IDUB) and the budget of the University of Warsaw (UW).

\end{document}